\PassOptionsToPackage{unicode}{hyperref}
\PassOptionsToPackage{hyphens}{url}
\PassOptionsToPackage{dvipsnames,svgnames,x11names}{xcolor}
\documentclass[
]{article}

\usepackage{amsmath,amssymb}
\usepackage{setspace}
\usepackage{iftex}
\ifPDFTeX
  \usepackage[T1]{fontenc}
  \usepackage[utf8]{inputenc}
  \usepackage{textcomp} 
\else 
  \usepackage{unicode-math}
  \defaultfontfeatures{Scale=MatchLowercase}
  \defaultfontfeatures[\rmfamily]{Ligatures=TeX,Scale=1}
\fi
\usepackage{lmodern}
\ifPDFTeX\else  
\fi
\IfFileExists{upquote.sty}{\usepackage{upquote}}{}
\IfFileExists{microtype.sty}{
  \usepackage[]{microtype}
  \UseMicrotypeSet[protrusion]{basicmath} 
}{}
\makeatletter
\@ifundefined{KOMAClassName}{
  \IfFileExists{parskip.sty}{%
    \usepackage{parskip}
  }{
    \setlength{\parindent}{0pt}
    \setlength{\parskip}{6pt plus 2pt minus 1pt}}
}{
  \KOMAoptions{parskip=half}}
\makeatother
\usepackage{xcolor}
\makeatletter
\ifx\paragraph\undefined\else
  \let\oldparagraph\paragraph
  \renewcommand{\paragraph}{
    \@ifstar
      \xxxParagraphStar
      \xxxParagraphNoStar
  }
  \newcommand{\xxxParagraphStar}[1]{\oldparagraph*{#1}\mbox{}}
  \newcommand{\xxxParagraphNoStar}[1]{\oldparagraph{#1}\mbox{}}
\fi
\ifx\subparagraph\undefined\else
  \let\oldsubparagraph\subparagraph
  \renewcommand{\subparagraph}{
    \@ifstar
      \xxxSubParagraphStar
      \xxxSubParagraphNoStar
  }
  \newcommand{\xxxSubParagraphStar}[1]{\oldsubparagraph*{#1}\mbox{}}
  \newcommand{\xxxSubParagraphNoStar}[1]{\oldsubparagraph{#1}\mbox{}}
\fi
\makeatother

\usepackage{color}
\usepackage{fancyvrb}

\DefineVerbatimEnvironment{Highlighting}{Verbatim}{commandchars=\\\{\}}
\usepackage{framed}
\definecolor{shadecolor}{RGB}{241,243,245}
\newenvironment{Shaded}{\begin{snugshade}}{\end{snugshade}}

\newcommand{\AttributeTok}[1]{\textcolor[rgb]{0.40,0.45,0.13}{#1}}

\newcommand{\DecValTok}[1]{\textcolor[rgb]{0.68,0.00,0.00}{#1}}

\newcommand{\FunctionTok}[1]{\textcolor[rgb]{0.28,0.35,0.67}{#1}}

\newcommand{\NormalTok}[1]{\textcolor[rgb]{0.00,0.23,0.31}{#1}}

\newcommand{\OtherTok}[1]{\textcolor[rgb]{0.00,0.23,0.31}{#1}}

\newcommand{\SpecialCharTok}[1]{\textcolor[rgb]{0.37,0.37,0.37}{#1}}

\providecommand{\tightlist}{%
  \setlength{\itemsep}{0pt}\setlength{\parskip}{0pt}}\usepackage{longtable,booktabs,array}
\usepackage{calc} 
\usepackage{etoolbox}
\makeatletter
\patchcmd\longtable{\par}{\if@noskipsec\mbox{}\fi\par}{}{}
\makeatother
\IfFileExists{footnotehyper.sty}{\usepackage{footnotehyper}}{\usepackage{footnote}}
\makesavenoteenv{longtable}
\usepackage{graphicx}
\makeatletter
\newsavebox\pandoc@box
\newcommand*\pandocbounded[1]{
  \sbox\pandoc@box{#1}%
  \Gscale@div\@tempa{\textheight}{\dimexpr\ht\pandoc@box+\dp\pandoc@box\relax}%
  \Gscale@div\@tempb{\linewidth}{\wd\pandoc@box}%
  \ifdim\@tempb\p@<\@tempa\p@\let\@tempa\@tempb\fi
  \ifdim\@tempa\p@<\p@\scalebox{\@tempa}{\usebox\pandoc@box}%
  \else\usebox{\pandoc@box}%
  \fi%
}
\def\fps@figure{htbp}
\makeatother
\NewDocumentCommand\citeproctext{}{}

\makeatletter
 \let\@cite@ofmt\@firstofone
 \def\@biblabel#1{}
 \def\@cite#1#2{{#1\if@tempswa , #2\fi}}
\makeatother
\newlength{\cslhangindent}
\newlength{\csllabelwidth}
\newenvironment{CSLReferences}[2] 
 {\begin{list}{}{%
  \setlength{\itemindent}{0pt}
  \setlength{\leftmargin}{0pt}
  \setlength{\parsep}{0pt}
  \ifodd #1
   \setlength{\leftmargin}{\cslhangindent}
   \setlength{\itemindent}{-1\cslhangindent}
  \fi
  \setlength{\itemsep}{#2\baselineskip}}}
 {\end{list}}
\usepackage{calc}

\newcommand{\CSLLeftMargin}[1]{\parbox[t]{\csllabelwidth}{\strut#1\strut}}
\newcommand{\CSLRightInline}[1]{\parbox[t]{\linewidth - \csllabelwidth}{\strut#1\strut}}

\usepackage{arxiv}
\usepackage{orcidlink}
\usepackage{amsmath}
\usepackage[T1]{fontenc}
\makeatletter
\@ifpackageloaded{caption}{}{\usepackage{caption}}
\AtBeginDocument{%
\ifdefined\contentsname
  \renewcommand*\contentsname{Table of contents}
\else
  \newcommand\contentsname{Table of contents}
\fi
\ifdefined\listfigurename
  \renewcommand*\listfigurename{List of Figures}
\else
  \newcommand\listfigurename{List of Figures}
\fi
\ifdefined\listtablename
  \renewcommand*\listtablename{List of Tables}
\else
  \newcommand\listtablename{List of Tables}
\fi
\ifdefined\figurename
  \renewcommand*\figurename{Figure}
\else
  \newcommand\figurename{Figure}
\fi
\ifdefined\tablename
  \renewcommand*\tablename{Table}
\else
  \newcommand\tablename{Table}
\fi
}
\@ifpackageloaded{float}{}{\usepackage{float}}
\floatstyle{ruled}
\@ifundefined{c@chapter}{\newfloat{codelisting}{h}{lop}}{\newfloat{codelisting}{h}{lop}[chapter]}
\floatname{codelisting}{Listing}

\usepackage{amsthm}
\theoremstyle{definition}
\newtheorem{example}{Example}[section]
\theoremstyle{remark}
\AtBeginDocument{}

\makeatother
\makeatletter
\@ifpackageloaded{caption}{}{\usepackage{caption}}
\@ifpackageloaded{subcaption}{}{\usepackage{subcaption}}
\makeatother

\usepackage{bookmark}

\IfFileExists{xurl.sty}{\usepackage{xurl}}{} 
\hypersetup{
  pdftitle={Using Mathlink Cubes to Introduce Data Wrangling with Examples in R},
  pdfauthor={Lucy D'Agostino McGowan},
  colorlinks=true,
  linkcolor={blue},
  filecolor={Maroon},
  citecolor={Blue},
  urlcolor={Blue},
  pdfcreator={LaTeX via pandoc}}

\newcommand{\runninghead}{A Preprint }
\title{Using Mathlink Cubes to Introduce Data Wrangling with Examples in
R}
\author{\textbf{Lucy D'Agostino McGowan}\\\\Wake Forest University,
Department of Statistical
Sciences\\Winston-Salem\\\href{mailto:mcgowald@wfu.edu}{mcgowald@wfu.edu}}
\date{}
\begin{document}
\maketitle
\begin{abstract}
\begin{quote}
This paper explores an innovative approach to teaching data wrangling
skills to students through hands-on activities before transitioning to
coding. Data wrangling, a critical aspect of data analysis, involves
cleaning, transforming, and restructuring data. We introduce the use of
a physical tool, mathlink cubes, to facilitate a tangible understanding
of data sets. This approach helps students grasp the concepts of data
wrangling before implementing them in coding languages such as R. We
detail a classroom activity that includes hands-on tasks paralleling
common data wrangling processes such as filtering, selecting, and
mutating, followed by their coding equivalents using R's \texttt{dplyr}
package.
\end{quote}
\end{abstract}

\setstretch{2}
\section{Introduction}\label{introduction}

Understanding how to properly handle and process data is a crucial skill
for students learning data analysis, one that can lead to more efficient
and accurate analyses (1,2). We refer to this process as \emph{data
wrangling}; it is also referred to as \emph{data munging}, \emph{data
manipulation}, \emph{data management}, among other names (2). This
involves cleaning, transforming, and restructuring data sets.

Working with data and wrangling tasks \emph{physically}, through tools
such as mathlink cubes, before transitioning to abstract representations
via coding can help enhance students' understanding of the material.
When targeting an abstract concept, such as how to manipulate a data set
too large to physically see, it can be helpful begin with tangible,
concrete illustrations (3). Physical materials, or \emph{manipulatives},
have been used to support a wide variety of education goals (4--6), and
mathematics specifically (7--11), for decades. A meta-analysis of the
efficacy of teaching mathematics with concrete manipulatives indicated
that the use of manipulatives has a positive impact on learning compared
to instruction with only abstract mathematical symbols, with the largest
impact on learning outcomes of retention (12). In computer science there
is a history of using physical manipulatives to teach programming,
dating back to 1995 with AlgoBlock (13), and research has shown that
manipulatives can be beneficial to student learning (14--16).
Additionally, the process of learning how to code can be intimidating
for beginners; moving away from the computer to engage in more
accessible, hands-on activities can significantly enhance the learning
experience (17). By using physical objects like mathlink cubes, students
can more easily grasp the concepts of data wrangling, as they can
visually and tangibly see the effects of their actions. This hands-on
approach can help students develop a deeper understanding of the data
wrangling process, which they can then translate to a more abstract
understanding needed to implement the wrangling techniques via code.
Notably, much of the research in this area is focused on primary and
secondary education (K-12); we are interested in assessing whether these
tools can be useful at the undergraduate level.

There are many coding languages that can be used for data wrangling; two
of the most prominent are SQL and R (18). SQL is a standard language for
managing and manipulating databases, widely used in industry.
Particularly through packages such as \texttt{dplyr} and \texttt{tidyr}
(19--21), (18) argue that R is capable of rivaling SQL in both
processing power and the elegance of its syntax. Likewise, once students
have a foundation of data wrangling in R, it is straightforward to
translate these skills to connect with database query languages such as
SQL (22,23). Additionally, there are packages such as \texttt{dbplyr}
which allow the user to write \texttt{dplyr} code in R and translate it
automatically into SQL (24). As (2) argue, the \texttt{dplyr} package in
R includes several simple functions that correspond in name to the most
common data wrangling tasks, which we will highlight further in
Section~\ref{sec-data-wrangle}; these tasks will serve as the foundation
for the hands-on activity that we will describe.

R community has previously supported the use of virtual manipulatives
for teaching data manipulation via TidyBlocks, a Scratch-like online
tool for learning data manipulation in R (25). This project, begun by
Maya Gans (26) and then maintained by Greg Wilson, had community
support, but is currently no longer under development (27).

In addition to hands-on activities helping to build a physical
connection to coding, integrating collaborative learning, where students
work in groups to solve data-related problems, can further enhance the
learning experience (28). Moreover, case-based learning has been shown
to enhance practical understanding and application skills (29).

This paper will be organized as follows. Section~\ref{sec-mathlink} will
introducing a physical representation of data sets using mathlink cubes.
Section~\ref{sec-data-wrangle} will then describe common data wrangling
tasks. Section~\ref{sec-classroom} will then introduce data wrangling
using mathlink cubes with clear extensions of these skills to coding
data wrangling tasks in R. The methods described here are intended for
undergraduate students in an introductory course designed to teach data
analysis skills. It is not expected that the students will have any
prior data analysis or computer science courses.

\section{Mathlink cubes to represent data sets}\label{sec-mathlink}

\begin{figure}

\centering{

\includegraphics[width=7in,height=\textheight,keepaspectratio]{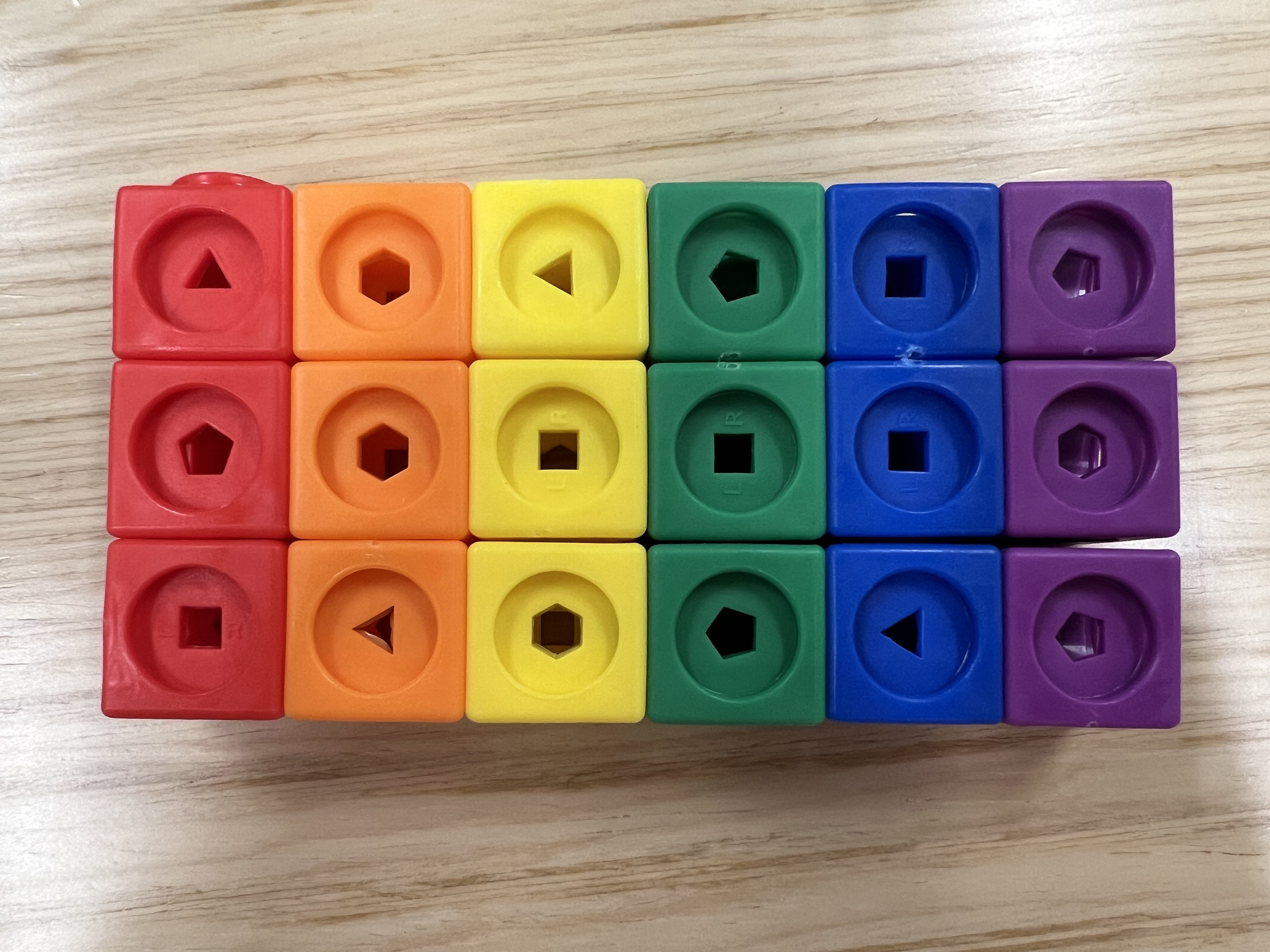}

}

\caption{\label{fig-1}Mathlink cubes arranged to represent a data set,
where each cube corresponds to a unique observation and the distinct
shapes on their faces, `triange', `square', `pentagon', and `hexagon',
denote variable values, 3, 4, 5, and 6, respectively. This `data set'
contains 6 variables, `red', `orange', `yellow', `green', `blue', and
`purple'.}

\end{figure}

Mathlink cubes, such as those created by Learning Resources (30), can be
used as a physical representation of a data set. The cubes can be
arranged in a grid formation, with each row representing an individual
observation and each column representing a variable
(Figure~\ref{fig-1}). The colors can indicate the variables, while the
shapes, visible on the faces of the cubes, indicate the values. For
example, a red cube with a triangle facing front can denote the value
`3' for the first variable in the first observation.

\begin{figure}

\centering{

\includegraphics[width=7in,height=\textheight,keepaspectratio]{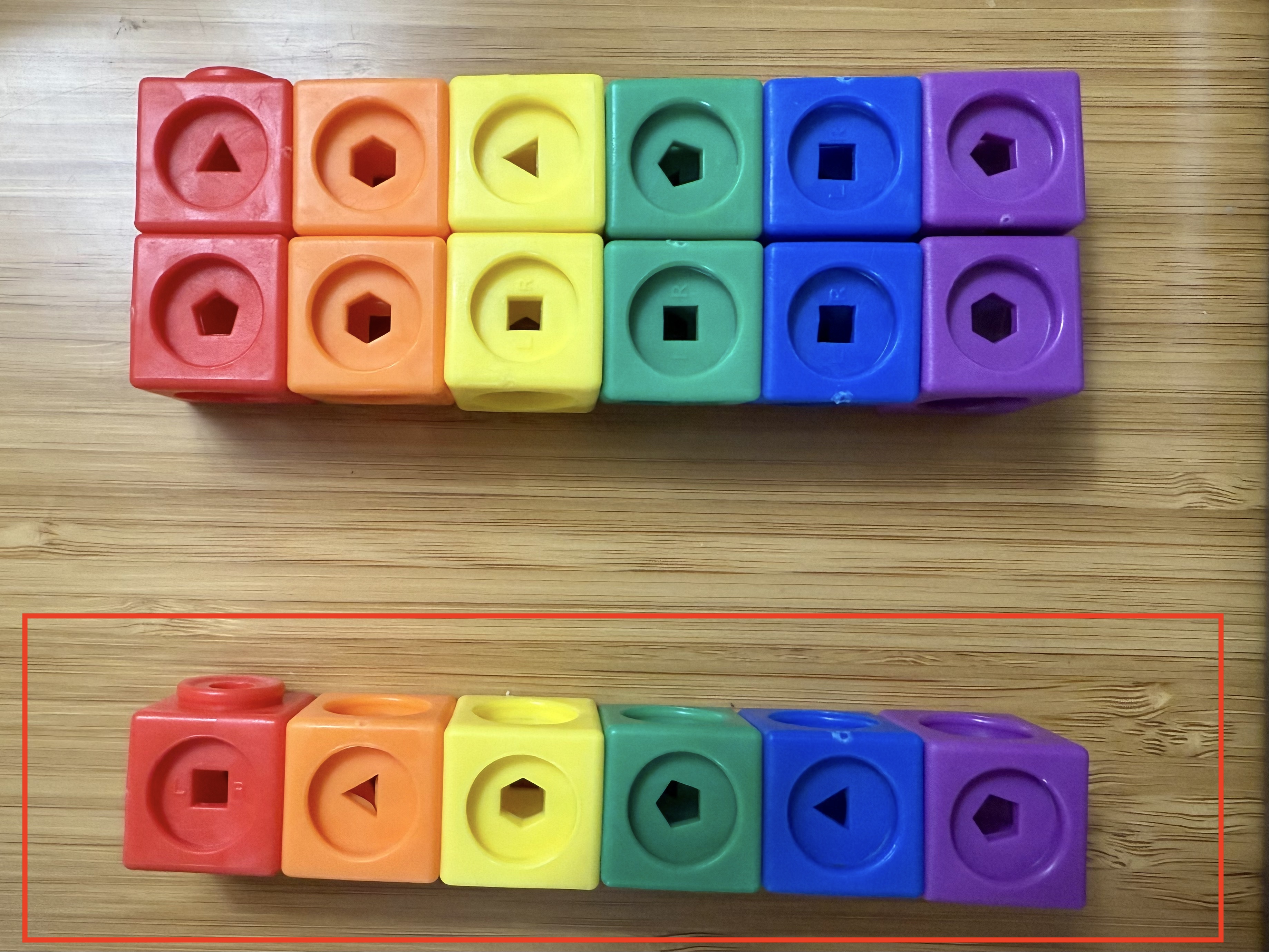}

}

\caption{\label{fig-2}Linear arrangement of mathlink cubes (appearing in
the red square), illustrating a single observation in the data set
represented in Figure~\ref{fig-1}, with variables differentiated by
color and values indicated by the shapes visible on the cube sides}

\end{figure}%

Mathlink cubes are designed to interlock both vertically (top to bottom)
and horizontally (side to side). A practical approach to assembling the
cube data set is to use the first column as an \emph{anchor}. In this
anchor column, the cubes are connected vertically, from top to bottom.
The subsequent columns are then attached horizontally to this anchor.
This arrangement not only provides stability but also enables individual
observations (rows) to be easily separated and examined
(Figure~\ref{fig-2}).

By handling these cubes, students can learn how to organize, sort, and
categorize data physically, mirroring the actions they will perform
abstractly when coding.

\section{Data wrangling tasks}\label{sec-data-wrangle}

Data wrangling involves a variety of tasks that clean, transform, and
restructure data into a format that is more suitable for analysis. Using
the same terminology as the \texttt{dplyr} package (19), some of the
most common tasks involved in data wrangling include filtering,
selecting, mutating, arranging, grouping, and summarizing, as described
below.

\textbf{Filtering} is the process of selecting rows that meet certain
logical conditions. This is analogous to extracting a subset of records
that share specific characteristics. For instance, an analyst might
filter a data set to include only observations where the variable age is
above a certain value so that an analysis can be completed only on this
subset of observations.

\textbf{\emph{Selecting}} is the process of choosing specific columns
that are relevant to the analysis. This involves focusing on particular
variables of interest and excluding others that are not pertinent to the
analysis question at hand.

\textbf{\emph{Mutating}} is the creation or transformation of variables
within a data set. This may include adding new variables that are
functions of existing ones. For example, an analyst might calculate the
body mass index (BMI) for a set of observations using information in two
other columns, height and weight, and add this as a new column to the
data set.

\textbf{\emph{Arranging}} is the process of sorting a data set by the
values within one or more columns. An analyst can order the observations
in ascending or descending order based on the values of the specified
variables.

\textbf{\emph{Grouping}} is the process of partitioning a data set into
subsets based on the unique values of one or more variables. Each group
consists of observations that share the same value for the grouping
variable(s), facilitating targeted analyses within these subsets. Often
grouping is used to facilitate the calculation of subsequent summary
statistics, for example, if an analyst wants to calculate the average
BMI within several age groups, they could first group the data by age.

\textbf{\emph{Summarizing}} involves the creation of a table of summary
statistics from a data set. For example, this could include calculating
means, medians, counts, or other aggregate metrics across a data set or,
if the data set is grouped, within each group. Summarizing can be used
to check the accuracy data wrangling of tasks, for example if a data set
contains a variable that has five categories and the analyst mutates the
data to created a new categorical variable that collapsed this into two
categories, they could summarize the data by counting the observations
in each unique combination of the old and new variable to ensure correct
re-categorization took place.

These common data wrangling tasks are supported by many programming
languages. In R, for example, the \texttt{dplyr} package offers a suite
of functions that correspond to each of these tasks.

\section{Classroom activity}\label{sec-classroom}

\subsection{Logistics}\label{logistics}

This activity was designed for small to medium class sizes (and tested
on a class of 30 students, see Section~\ref{sec-data}) to be completed
in a classroom with either desks or tables for the students to complete
the activity on. Scalability of the activity will depend on two factors:
(1) the number of mathlink cubes needed to complete the activity and (2)
the amount of time needed to get the activity started (i.e.~pass out the
cubes). Prior to the classroom activity, the instructor can create 3x6
dimensional mathlink data sets as shown in Figure~\ref{fig-1} as well as
one set of 6 additional mathlinks cubes in a 7th color that can be used
for the summary portion. The number of mathlink data sets to create
depends on the number of students in the class; we created enough so
that students could work in groups of 3. We had one student from each
group come forward at the beginning of the activity to collect the cubes
they would need to complete the activities. The hands-on activities were
completed during one 75-minute class period. Example slides are included
in the Supplemental Materials.

\subsection{Hands-on activity}\label{sec-hands-on}

The hands-on activity begins by introducing the students to their
mathlink data set. Example questions to ask the students after they have
their mathlink data sets in hand are:

\begin{example}[]\protect\hypertarget{exm-questions}{}\label{exm-questions}

~

\begin{enumerate}
\def\labelenumi{(\arabic{enumi})}
\tightlist
\item
  How many observations are in your data set? (Answer: 3)
\item
  What are the unique values for observations in this data set? (Answer:
  3, 4, 5, 6, corresponding to the ``triangle'', ``square'',
  ``pentagon'' and ``hexagon'' shapes)
\item
  How many columns are in your data set? (Answer: 6)
\item
  What are names would you give the columns? (Answer: ``red'', orange,
  ``yellow'', ``green'', ``blue'', ``purple'')
\end{enumerate}

\end{example}

We then introduce each of the data wrangling tasks described in
Section~\ref{sec-data-wrangle}, followed by tasks that allow the
students to manipulate their mathlink data sets.

\subsubsection{Filter}\label{filter}

We begin with \emph{filtering}. For this first task, we start with a
description of the task in plain English:

\begin{example}[]\protect\hypertarget{exm-longhand}{}\label{exm-longhand}

\textbf{Take your data frame and then filter it} (only include rows
where) \textbf{the red column only includes observations with three
sides} (triangles) \textbf{OR the green column only includes
observations with more than 4 sides} (pentagons, hexagons)

\end{example}

\begin{figure}

\begin{minipage}{0.50\linewidth}

\centering{

\includegraphics[width=0.95\linewidth,height=\textheight,keepaspectratio]{figs/fig1.jpg}

}

\subcaption{\label{fig-3-1}Original arrangement}

\end{minipage}%
\begin{minipage}{0.50\linewidth}

\centering{

\includegraphics[width=0.95\linewidth,height=\textheight,keepaspectratio]{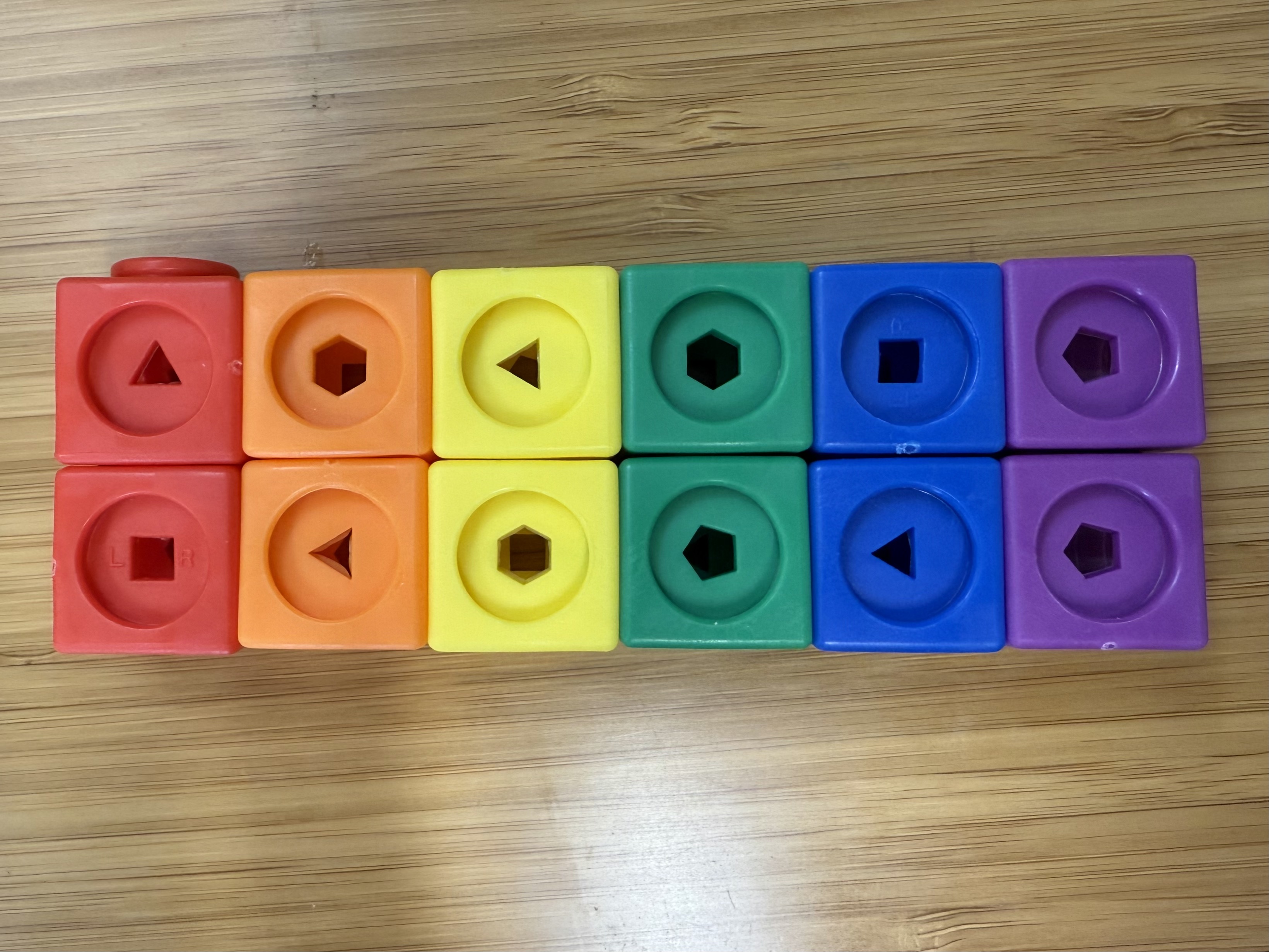}

}

\subcaption{\label{fig-3-2}Filtered arrangement}

\end{minipage}%

\caption{\label{fig-3}Mathlink cubes filtered according to
Example~\ref{exm-longhand}}

\end{figure}

After students complete this task in their groups, the instructor can
ask at least two groups to demonstrate how they completed the task and
show what their final, filtered, data set looks like. We can then
discuss the dimensions of the filtered data set. Depending on how their
initial mathlink data sets were arranged, groups may have different
answers. Figure~\ref{fig-3} shows how a mathlink data set arranged as in
Figure~\ref{fig-1} would be correctly filtered according to
Example~\ref{exm-longhand}.

A common pitfall students encounter here is many may both remove rows
\emph{and} columns (i.e.~result in a final data set with only the red
and green columns, as those were the targets of this task). This is a
great opportunity to remind students that filtering will never change
the dimensions of the columns, it will only (potentially) reduce the
number of rows. Another common pitfall is students will erroneously try
to join the two logical statements with an \emph{and} operator rather
than \emph{or}. This is a good opportunity to remind students of Boolean
operators and what they mean.

Next, we introduce logical and Boolean operator short hand, for example
using the information in Table~\ref{tbl-logical}.

\begin{longtable}[]{@{}ll@{}}
\caption{Logical and Boolean shorthand translated to plain
language}\label{tbl-logical}\tabularnewline
\toprule\noalign{}
\textbf{Short hand} & \textbf{Meaning} \\
\midrule\noalign{}
\endfirsthead
\toprule\noalign{}
\textbf{Short hand} & \textbf{Meaning} \\
\midrule\noalign{}
\endhead
\bottomrule\noalign{}
\endlastfoot
x \textless{} y & x is less than y \\
x \textgreater{} y & x is greater than y \\
x \textless= y & x is less than or equal to y \\
x \textgreater= y & x is greater than or equal to y \\
x == y & x is exactly equal to y \\
x != y & x is not equal to y \\
x \%in\% y & x is contained in y \\
is.na(x) & x is missing \\
!is.na(x) & x is not missing \\
a \& b & a and b \\
a \textbar{} b & a or b \\
!a & not a \\
\end{longtable}

The instructor can proceed by explaining that rather than having a wall
of text, as in Example~\ref{exm-longhand}, to describe each of the
tasks, we will use shorthand:

\begin{example}[]\protect\hypertarget{exm-shorthand}{}\label{exm-shorthand}

~

\begin{Shaded}
\begin{Highlighting}[]
\NormalTok{data }\SpecialCharTok{|\textgreater{}}
  \FunctionTok{filter}\NormalTok{(red }\SpecialCharTok{==} \DecValTok{3} \SpecialCharTok{|} 
\NormalTok{         green }\SpecialCharTok{\textgreater{}} \DecValTok{4}\NormalTok{)}
\end{Highlighting}
\end{Shaded}

\end{example}

If the instructor is using slides, they can step through each component
of Example~\ref{exm-shorthand} and map them back to the plain language
in Example~\ref{exm-longhand}, i.e.~\texttt{data}: ``take your data
frame'', \texttt{\textbar{}\textgreater{}} ``and then'',
\texttt{filter}: ``filter it (only include rows where)'',
\texttt{red\ ==\ 3}: ``the red column only includes observations with
three sides (triangles)'', \texttt{\textbar{}}: ``OR'',
\texttt{green\ \textgreater{}\ 4}: ``the green column only includes
observations with more than 4 sides (pentagons, hexagons)''. Example
slides are included in the Supplemental Material. The instructor can
proceed using this shorthand (which happens to match the R syntax we
will teach in a subsequent lesson) to request more filtering tasks,
asking different groups each time to describe their solution.

\subsubsection{Select}\label{select}

\begin{figure}

\begin{minipage}{\linewidth}

\centering{

\includegraphics[width=0.9\linewidth,height=\textheight,keepaspectratio]{figs/fig1.jpg}

}

\subcaption{\label{fig-4-1}Original arrangement}

\end{minipage}%
\newline
\begin{minipage}{0.45\linewidth}

\centering{

\includegraphics[width=0.9\linewidth,height=\textheight,keepaspectratio]{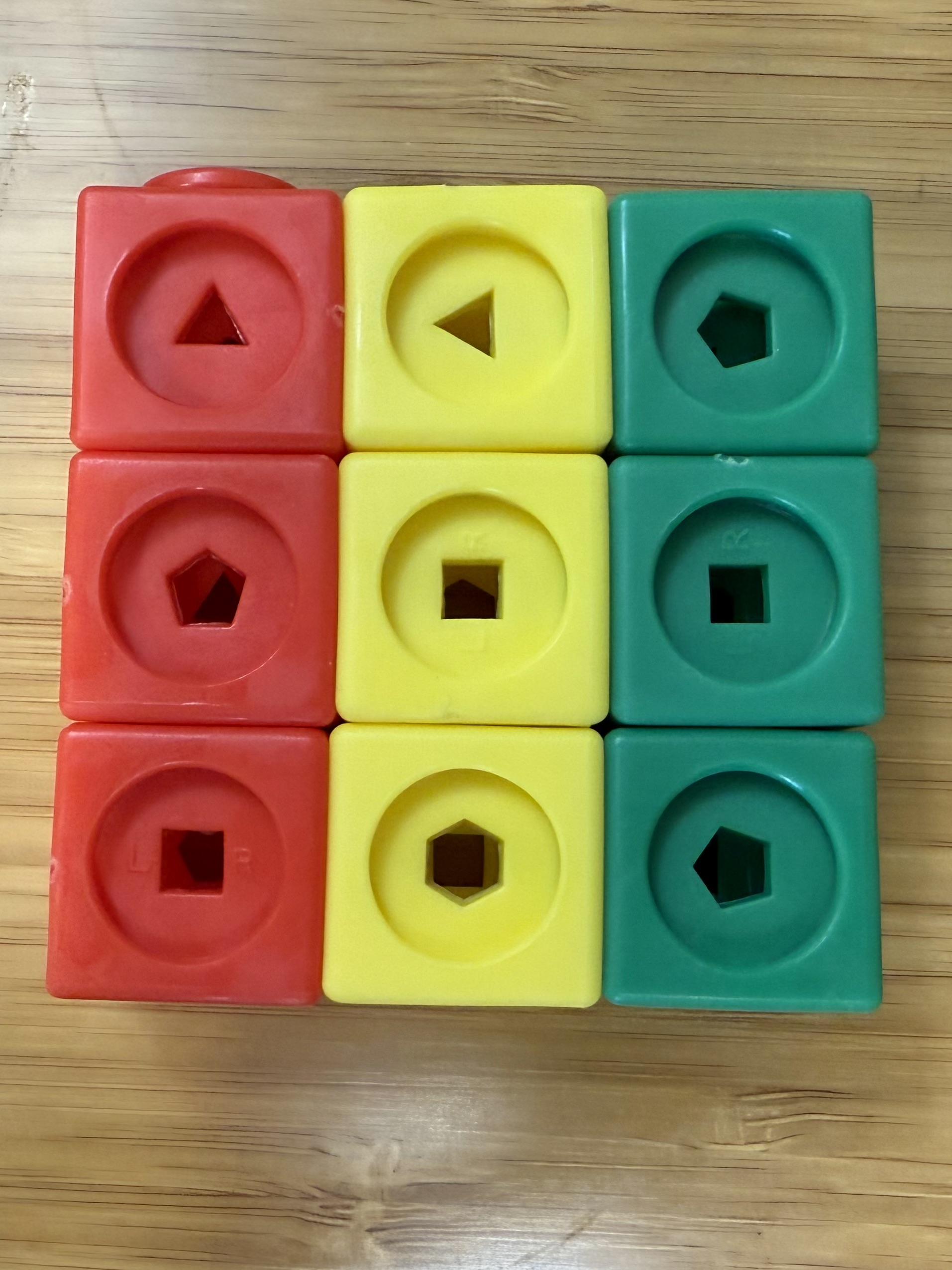}

}

\subcaption{\label{fig-4-2}Select according to Example~\ref{exm-select}
A.}

\end{minipage}%
\begin{minipage}{0.10\linewidth}
~\end{minipage}%
\begin{minipage}{0.45\linewidth}

\centering{

\includegraphics[width=0.9\linewidth,height=\textheight,keepaspectratio]{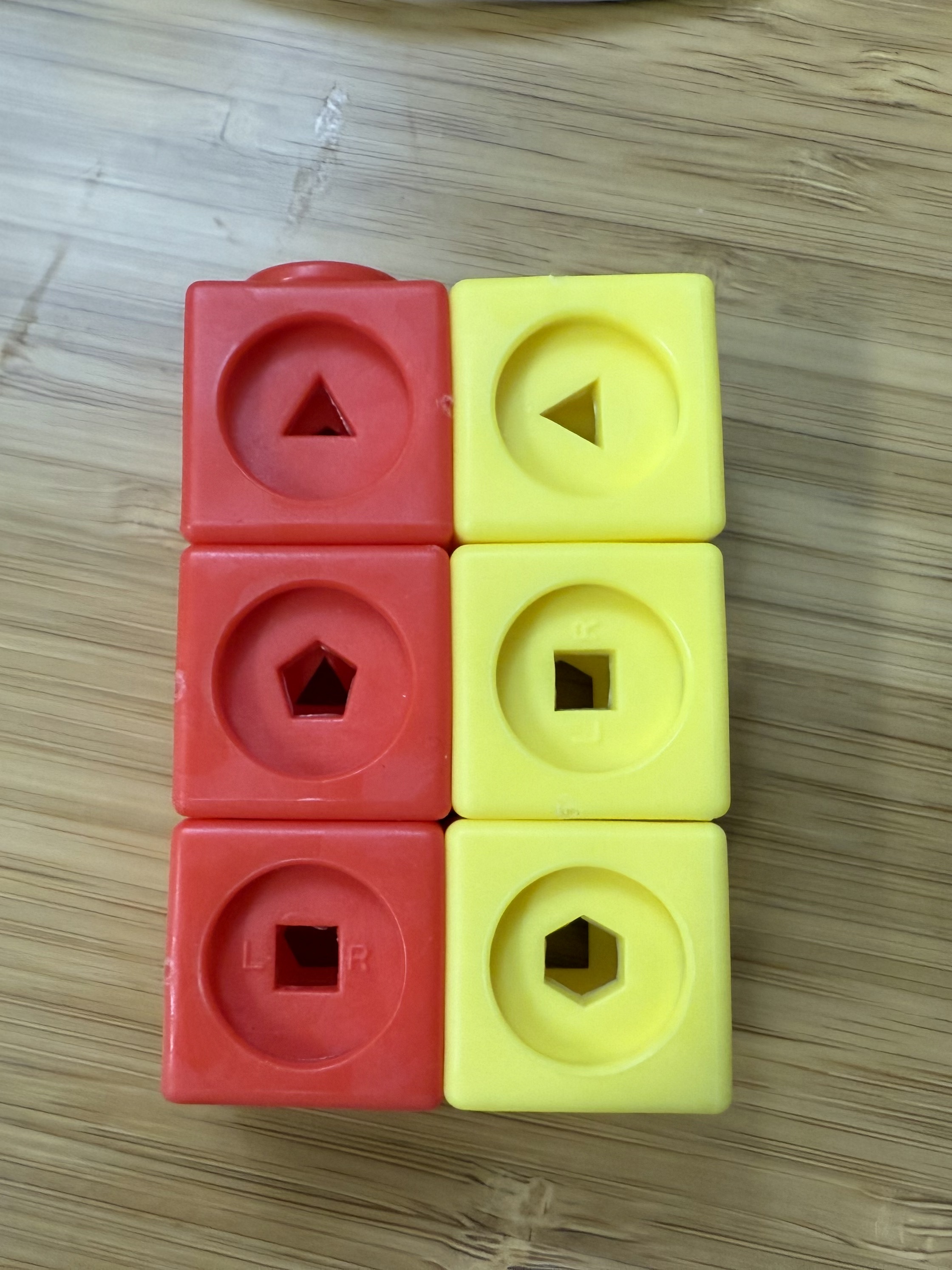}

}

\subcaption{\label{fig-4-3}Select again according to
Example~\ref{exm-select} B.}

\end{minipage}%

\caption{\label{fig-4}Mathlink cubes selected according to
Example~\ref{exm-select}}

\end{figure}

We then proceed to the \emph{select} task. The instructor should begin
by asking students to return the three observations to the data set so
they begin with a 3x6 data set again. After explaining the concept, we
ask students to complete selection tasks using their mathlink data sets,
such as those shown in Example~\ref{exm-select}. These can show both
selecting columns by name (Example~\ref{exm-select} A) and also removing
columns by name (Example~\ref{exm-select} B) (Figure~\ref{fig-4}). We
can also introduce here that if we want to start where we left off
(i.e., start with Example~\ref{exm-select} A and then remove one more
column in Example~\ref{exm-select} B), we need to explictly save our
work (in this case, by assigning the code in Example~\ref{exm-select} A
to the object \texttt{data\_a}).

\pagebreak

\begin{example}[]\protect\hypertarget{exm-select}{}\label{exm-select}

~

\begin{figure}

\begin{minipage}{0.50\linewidth}

\begin{enumerate}
\def\labelenumi{\Alph{enumi}.}
\tightlist
\item
\end{enumerate}

\begin{Shaded}
\begin{Highlighting}[]
\NormalTok{data }\SpecialCharTok{|\textgreater{}}
  \FunctionTok{select}\NormalTok{(red, yellow, green)}
\end{Highlighting}
\end{Shaded}

\end{minipage}%
\begin{minipage}{0.50\linewidth}

\begin{enumerate}
\def\labelenumi{\Alph{enumi}.}
\setcounter{enumi}{1}
\tightlist
\item
\end{enumerate}

\begin{Shaded}
\begin{Highlighting}[]
\NormalTok{data\_a }\OtherTok{\textless{}{-}}\NormalTok{ data }\SpecialCharTok{|\textgreater{}}
  \FunctionTok{select}\NormalTok{(red, yellow, green)}
\NormalTok{data\_a }\SpecialCharTok{|\textgreater{}}
  \FunctionTok{select}\NormalTok{(}\SpecialCharTok{{-}}\NormalTok{green)}
\end{Highlighting}
\end{Shaded}

\end{minipage}%

\end{figure}%

\end{example}

\subsubsection{Mutate}\label{mutate}

\begin{figure}

\begin{minipage}{0.50\linewidth}

\centering{

\includegraphics[width=0.95\linewidth,height=\textheight,keepaspectratio]{figs/fig5.jpeg}

}

\subcaption{\label{fig-5-1}Original arrangement}

\end{minipage}%
\begin{minipage}{0.50\linewidth}

\centering{

\includegraphics[width=0.95\linewidth,height=\textheight,keepaspectratio]{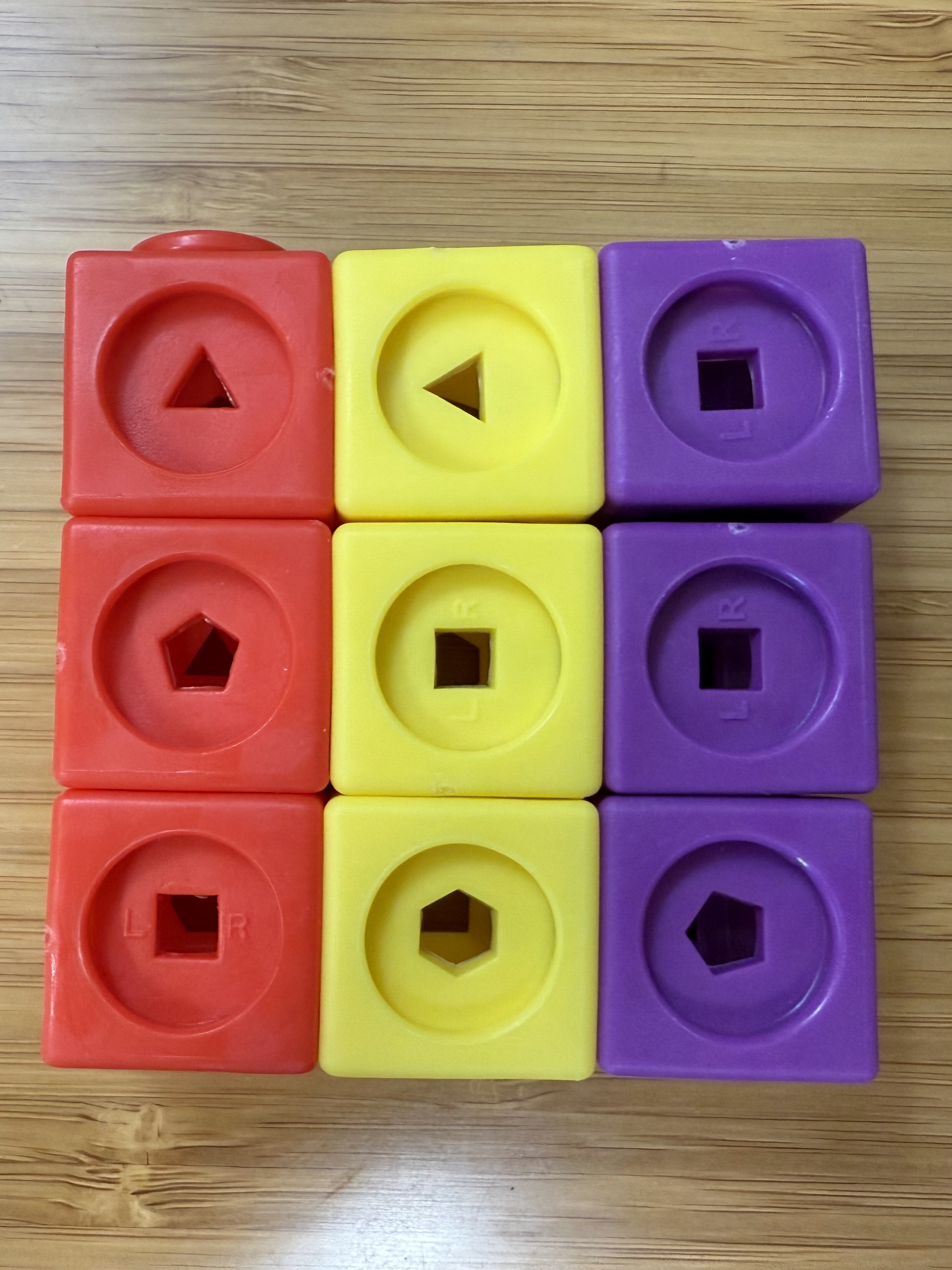}

}

\subcaption{\label{fig-5-2}Mutate according to
Example~\ref{exm-purple}.}

\end{minipage}%

\caption{\label{fig-5}Mathlink cubes mutated according to
Example~\ref{exm-purple}}

\end{figure}

We then introduce the \emph{mutate} task. Since we previously removed
variables using the \emph{select} we can add these same variables back
in. Begin by adding in a variable directly, not incorporating
information from other columns, for example Example~\ref{exm-purple}
adds the purple column in with values \texttt{4}, \texttt{4},
\texttt{5}, corresponding to the first three observations (meaning the
student's final column should consist of the purple mathlink cubes with
the ``square'', ``square'', ``pentagon'' facing forward,
Figure~\ref{fig-5}).

\begin{example}[]\protect\hypertarget{exm-purple}{}\label{exm-purple}

~

\begin{Shaded}
\begin{Highlighting}[]
\NormalTok{data }\SpecialCharTok{|\textgreater{}}
  \FunctionTok{mutate}\NormalTok{(}\AttributeTok{purple =} \FunctionTok{c}\NormalTok{(}\DecValTok{4}\NormalTok{, }\DecValTok{4}\NormalTok{, }\DecValTok{5}\NormalTok{))}
\end{Highlighting}
\end{Shaded}

\end{example}

After having the students add a few variables directly, verbs that take
logical statements into account can be introduced (such as
``if\ldots else'' or ``case when''). For example, beginning with a plain
language explanation, ``if\ldots else'' can be explained as:

\begin{example}[]\protect\hypertarget{exm-ifelse}{}\label{exm-ifelse}

\textbf{If} a logical statement is true, output the following, or
\textbf{else} output something else.

\end{example}

Example~\ref{exm-ifelse} can then be mapped to short-hand as follows:

\begin{example}[]\protect\hypertarget{exm-ifelse-short}{}\label{exm-ifelse-short}

~

\begin{Shaded}
\begin{Highlighting}[]
\FunctionTok{ifelse}\NormalTok{(logical\_test, }
\NormalTok{       value\_if\_true,}
\NormalTok{       value\_if\_false)}
\end{Highlighting}
\end{Shaded}

\end{example}

After showing Example~\ref{exm-ifelse-short}, it might be a good
opportunity to remind students of the logical statement shorthand, such
as those shown in Table~\ref{tbl-logical}. We can then move on to
demonstrating how to mutate a data set to build new variables using
existing columns, as shown in Example~\ref{exm-mutate}.

\newpage

\begin{example}[]\protect\hypertarget{exm-mutate}{}\label{exm-mutate}

~

\begin{figure}

\begin{minipage}{0.50\linewidth}

\begin{enumerate}
\def\labelenumi{\Alph{enumi}.}
\tightlist
\item
\end{enumerate}

\begin{Shaded}
\begin{Highlighting}[]
\NormalTok{data }\SpecialCharTok{|\textgreater{}}
  \FunctionTok{mutate}\NormalTok{(}
    \AttributeTok{blue =} \FunctionTok{ifelse}\NormalTok{(red }\SpecialCharTok{\textgreater{}} \DecValTok{3}\NormalTok{, }\DecValTok{4}\NormalTok{, }\DecValTok{5}\NormalTok{)}
\NormalTok{  )}
\end{Highlighting}
\end{Shaded}

\end{minipage}%
\begin{minipage}{0.50\linewidth}

\begin{enumerate}
\def\labelenumi{\Alph{enumi}.}
\setcounter{enumi}{1}
\tightlist
\item
\end{enumerate}

\begin{Shaded}
\begin{Highlighting}[]
\NormalTok{data }\SpecialCharTok{|\textgreater{}}
  \FunctionTok{mutate}\NormalTok{(}
    \AttributeTok{orange =} \FunctionTok{ifelse}\NormalTok{(blue }\SpecialCharTok{==} \DecValTok{6}\NormalTok{, }\DecValTok{4}\NormalTok{, }\DecValTok{3}\NormalTok{),}
    \AttributeTok{green =}\NormalTok{ orange }\SpecialCharTok{+} \DecValTok{1}
\NormalTok{  )}
\end{Highlighting}
\end{Shaded}

\end{minipage}%

\end{figure}%

\end{example}

\subsubsection{Arrange}\label{arrange}

\begin{figure}

\begin{minipage}{0.50\linewidth}

\centering{

\includegraphics[width=0.95\linewidth,height=\textheight,keepaspectratio]{figs/fig6.jpeg}

}

\subcaption{\label{fig-6-1}Original arrangement}

\end{minipage}%
\begin{minipage}{0.50\linewidth}

\centering{

\includegraphics[width=0.95\linewidth,height=\textheight,keepaspectratio]{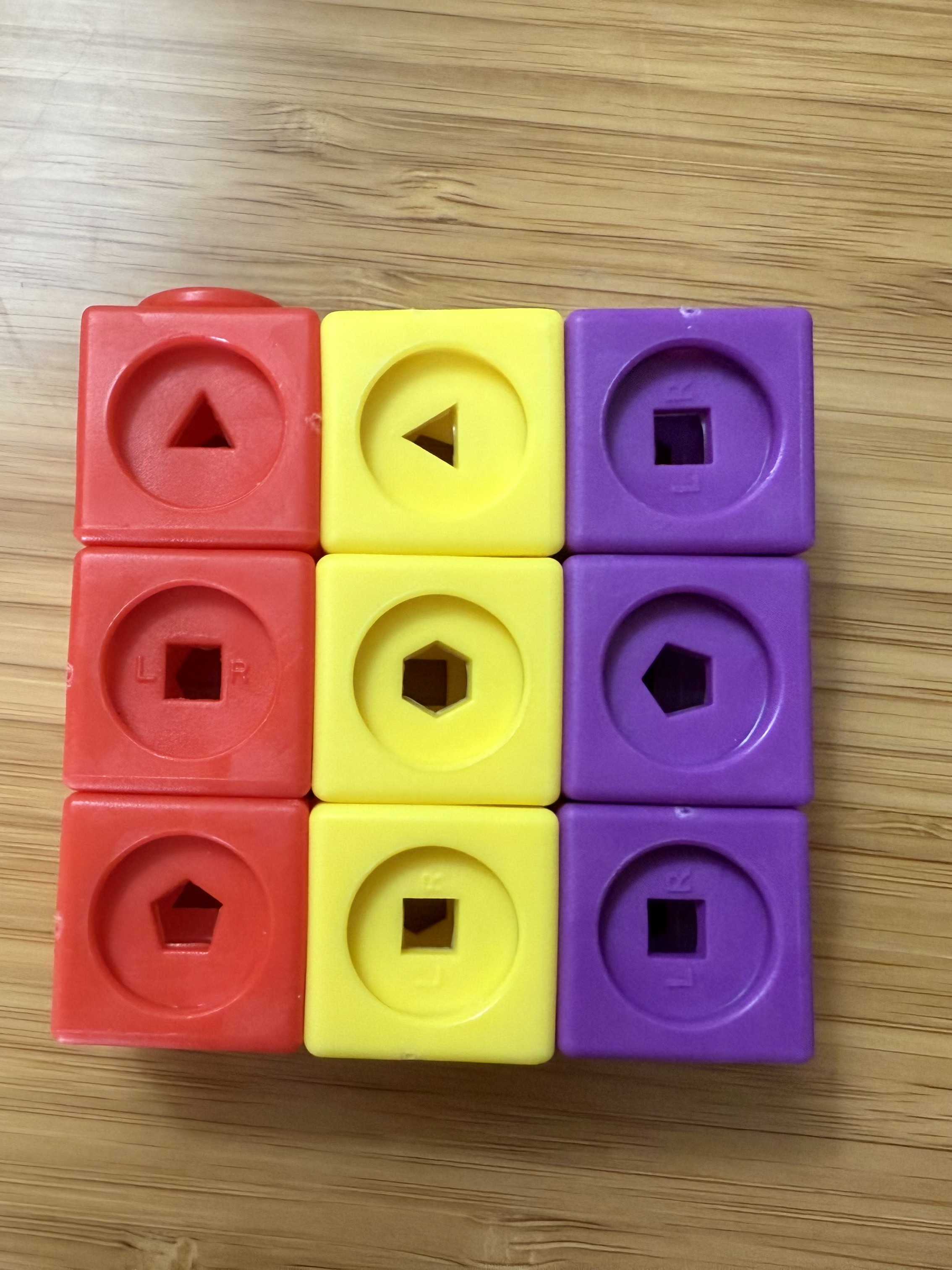}

}

\subcaption{\label{fig-6-2}Arrange according to the red column in
ascending order, as in Example~\ref{exm-arrange} A.}

\end{minipage}%

\caption{\label{fig-6}Mathlink cubes arranged according to
Example~\ref{exm-arrange} A.}

\end{figure}

Next, we demonstrate how to \emph{arrange} a data set, asking the
students to sort their data based on the values of specified columns. By
default, we will arrange in \emph{ascending} order, but this could be
reversed by indicating that descending order is preferred, using the
shorthand \texttt{desc} (Example~\ref{exm-arrange}, Figure~\ref{fig-6}).

\begin{example}[]\protect\hypertarget{exm-arrange}{}\label{exm-arrange}

~

\begin{figure}

\begin{minipage}{0.50\linewidth}

\begin{enumerate}
\def\labelenumi{\Alph{enumi}.}
\tightlist
\item
\end{enumerate}

\begin{Shaded}
\begin{Highlighting}[]
\NormalTok{data }\SpecialCharTok{|\textgreater{}}
  \FunctionTok{arrange}\NormalTok{(red)}
\end{Highlighting}
\end{Shaded}

\end{minipage}%
\begin{minipage}{0.50\linewidth}

\begin{enumerate}
\def\labelenumi{\Alph{enumi}.}
\setcounter{enumi}{1}
\tightlist
\item
\end{enumerate}

\begin{Shaded}
\begin{Highlighting}[]
\NormalTok{data }\SpecialCharTok{|\textgreater{}}
  \FunctionTok{arrange}\NormalTok{(}\FunctionTok{desc}\NormalTok{(red))}
\end{Highlighting}
\end{Shaded}

\end{minipage}%

\end{figure}%

\end{example}

\pagebreak

\subsubsection{Grouping}\label{grouping}

\begin{figure}

\begin{minipage}{0.45\linewidth}

\centering{

\includegraphics[width=0.95\linewidth,height=\textheight,keepaspectratio]{figs/fig7.jpeg}

}

\subcaption{\label{fig-7-1}Original arrangement and
Example~\ref{exm-groupby} A., as grouping does not change how the data
looks.}

\end{minipage}%
\begin{minipage}{0.10\linewidth}
~\end{minipage}%
\begin{minipage}{0.45\linewidth}

\centering{

\includegraphics[width=0.95\linewidth,height=\textheight,keepaspectratio]{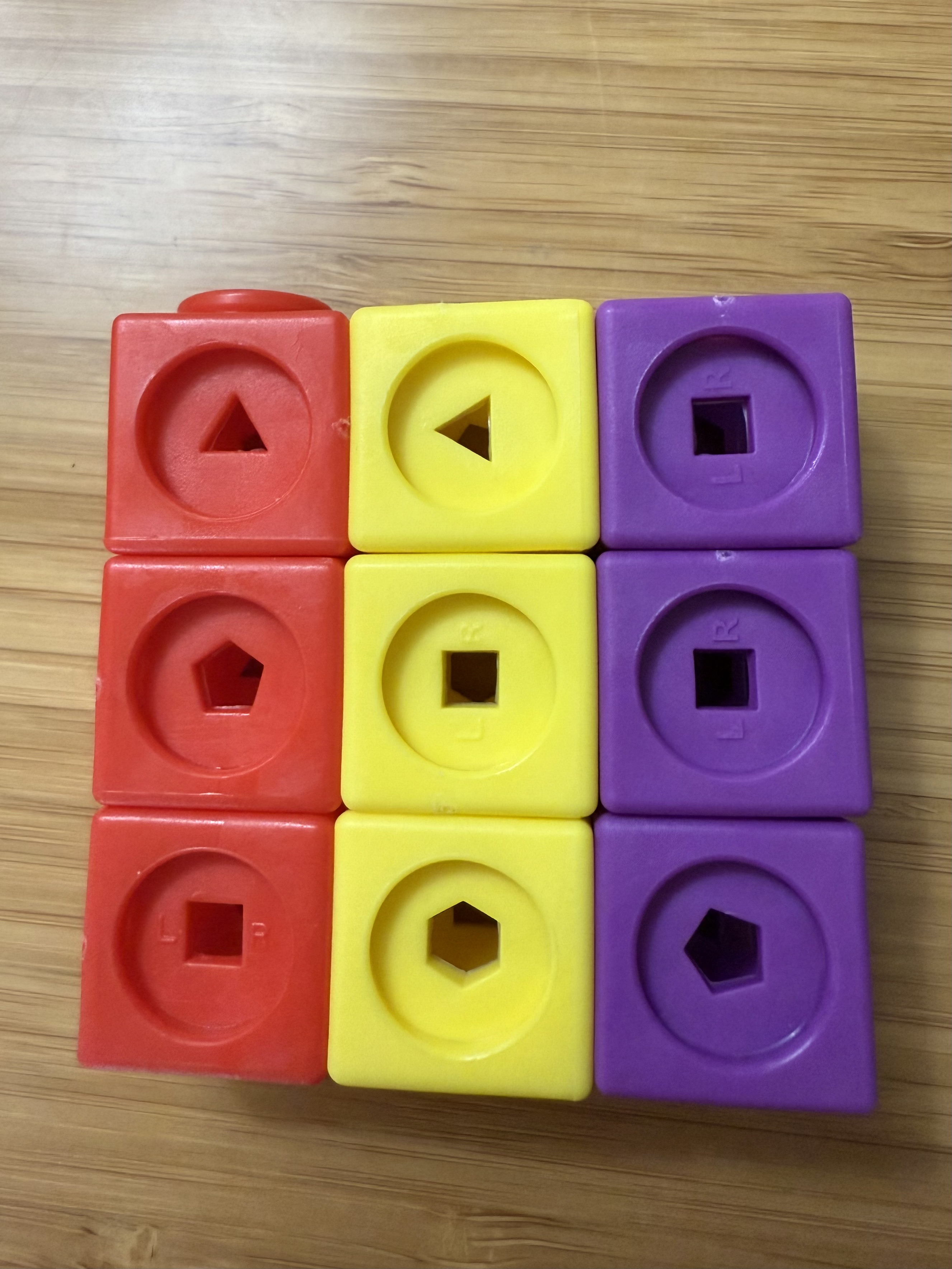}

}

\subcaption{\label{fig-7-2}Group by purple and arranged by red,
according to Example~\ref{exm-groupby}.}

\end{minipage}%

\caption{\label{fig-7}Mathlink cubes grouped and arranged according to
Example~\ref{exm-groupby}. Note that (b) has the cubes grouped by
purple, with the two purple squares in the top rows and the pentagon
beneath, and then within each purple grouping, the rows are arranged by
red (e.g., 3, 4 for the purple squares).}

\end{figure}

Then we demonstrate how to partition the data set into groups based on
values in the columns. We can start by showing how grouping does not
actually change the data set, it just registers what the groups are,
then we can show how it can be combined with another task like arranging
(Example~\ref{exm-groupby}, Figure~\ref{fig-7}).

\begin{example}[]\protect\hypertarget{exm-groupby}{}\label{exm-groupby}

~

\begin{figure}

\begin{minipage}{0.50\linewidth}

\begin{enumerate}
\def\labelenumi{\Alph{enumi}.}
\tightlist
\item
\end{enumerate}

\begin{Shaded}
\begin{Highlighting}[]
\NormalTok{data }\SpecialCharTok{|\textgreater{}}
  \FunctionTok{group\_by}\NormalTok{(purple)}
\end{Highlighting}
\end{Shaded}

\end{minipage}%
\begin{minipage}{0.50\linewidth}

\begin{enumerate}
\def\labelenumi{\Alph{enumi}.}
\setcounter{enumi}{1}
\tightlist
\item
\end{enumerate}

\begin{Shaded}
\begin{Highlighting}[]
\NormalTok{data }\SpecialCharTok{|\textgreater{}}
  \FunctionTok{group\_by}\NormalTok{(purple) }\SpecialCharTok{|\textgreater{}}
  \FunctionTok{arrange}\NormalTok{(red)}
\end{Highlighting}
\end{Shaded}

\end{minipage}%

\end{figure}%

\end{example}

\pagebreak

\subsubsection{Summarize}\label{summarize}

Next, we can introduce \emph{summarizing} the data set. For this, a
separate set of mathlink cubes in a different color than the six in the
data set are provided. A table of short hand for common summary
functions can be provided, such as shown in Table~\ref{tbl-2}.

\begin{longtable}[]{@{}ll@{}}
\caption{Common summary functions}\label{tbl-2}\tabularnewline
\toprule\noalign{}
\textbf{functions} & \textbf{Meaning} \\
\midrule\noalign{}
\endfirsthead
\toprule\noalign{}
\textbf{functions} & \textbf{Meaning} \\
\midrule\noalign{}
\endhead
\bottomrule\noalign{}
\endlastfoot
\texttt{min()} & minimum \\
\texttt{max()} & maximum \\
\texttt{mean()} & average \\
\texttt{sd()} & standard deviation \\
\texttt{sum()} & addition \\
\texttt{quantile(x,\ probs\ =\ 0.25)} & quantile, set probs {[}0 -
1{]} \\
\end{longtable}

We begin by introducing summary tasks that will create a single row of
summary statistics, as seen in the left example in
Example~\ref{exm-summary}. Then we can extend this to multiple rows by
integrating grouping tasks, as seen on the right.

\begin{example}[]\protect\hypertarget{exm-summary}{}\label{exm-summary}

~

\begin{figure}

\begin{minipage}{0.50\linewidth}

\begin{enumerate}
\def\labelenumi{\Alph{enumi}.}
\tightlist
\item
\end{enumerate}

\begin{Shaded}
\begin{Highlighting}[]
\NormalTok{data }\SpecialCharTok{|\textgreater{}}
  \FunctionTok{summarize}\NormalTok{(}
    \FunctionTok{max}\NormalTok{(red),}
    \FunctionTok{max}\NormalTok{(blue),}
    \FunctionTok{min}\NormalTok{(orange)}
\NormalTok{  )}
\end{Highlighting}
\end{Shaded}

\end{minipage}%
\begin{minipage}{0.50\linewidth}

\begin{enumerate}
\def\labelenumi{\Alph{enumi}.}
\setcounter{enumi}{1}
\tightlist
\item
\end{enumerate}

\begin{Shaded}
\begin{Highlighting}[]
\NormalTok{data }\SpecialCharTok{|\textgreater{}}
  \FunctionTok{group\_by}\NormalTok{(blue) }\SpecialCharTok{|\textgreater{}}
  \FunctionTok{summarize}\NormalTok{(}
    \FunctionTok{min}\NormalTok{(red),}
    \FunctionTok{max}\NormalTok{(green)}
\NormalTok{  )}
\end{Highlighting}
\end{Shaded}

\end{minipage}%

\end{figure}%

\end{example}

\subsubsection{Combining tasks}\label{combining-tasks}

Finally, we can demonstrate that the previous tasks can be combined, for
example asking the students to complete the tasks shown in
Example~\ref{exm-combined}.

\begin{example}[]\protect\hypertarget{exm-combined}{}\label{exm-combined}

~

\begin{figure}

\begin{minipage}{0.50\linewidth}

\begin{enumerate}
\def\labelenumi{\Alph{enumi}.}
\tightlist
\item
\end{enumerate}

\begin{Shaded}
\begin{Highlighting}[]
\NormalTok{data }\SpecialCharTok{|\textgreater{}}
  \FunctionTok{filter}\NormalTok{(blue }\SpecialCharTok{\textgreater{}} \DecValTok{3}\NormalTok{) }\SpecialCharTok{|\textgreater{}}
  \FunctionTok{select}\NormalTok{(red, yellow, blue)}\SpecialCharTok{|\textgreater{}}
  \FunctionTok{mutate}\NormalTok{(}\AttributeTok{green =}\NormalTok{ blue }\SpecialCharTok{{-}} \DecValTok{1}\NormalTok{)}
\end{Highlighting}
\end{Shaded}

\end{minipage}%
\begin{minipage}{0.50\linewidth}

\begin{enumerate}
\def\labelenumi{\Alph{enumi}.}
\setcounter{enumi}{1}
\tightlist
\item
\end{enumerate}

\begin{Shaded}
\begin{Highlighting}[]
\NormalTok{data }\SpecialCharTok{|\textgreater{}}
  \FunctionTok{filter}\NormalTok{(blue }\SpecialCharTok{\textgreater{}} \DecValTok{4}\NormalTok{) }\SpecialCharTok{|\textgreater{}}
  \FunctionTok{summarize}\NormalTok{(}\FunctionTok{max}\NormalTok{(blue))}
\end{Highlighting}
\end{Shaded}

\end{minipage}%

\end{figure}%

\end{example}

\begin{figure}

\begin{minipage}{\linewidth}

\centering{

\includegraphics[width=0.9\linewidth,height=\textheight,keepaspectratio]{figs/fig1.jpg}

}

\subcaption{\label{fig-8-1}Original arrangement}

\end{minipage}%
\newline
\begin{minipage}{0.45\linewidth}

\centering{

\includegraphics[width=0.9\linewidth,height=\textheight,keepaspectratio]{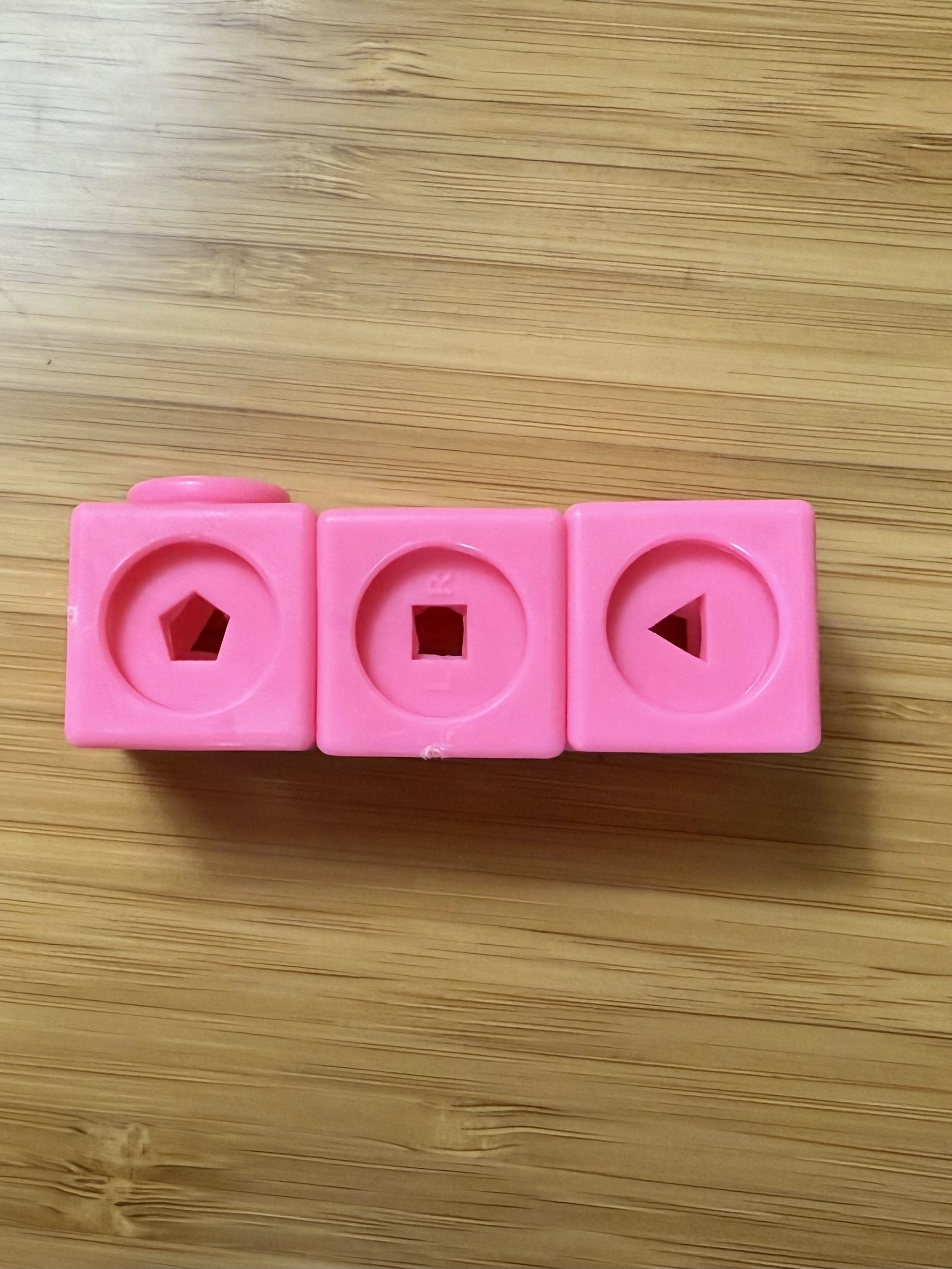}

}

\subcaption{\label{fig-8-2}Summarize according to
Example~\ref{exm-summary} A.}

\end{minipage}%
\begin{minipage}{0.10\linewidth}
~\end{minipage}%
\begin{minipage}{0.45\linewidth}

\centering{

\includegraphics[width=0.9\linewidth,height=\textheight,keepaspectratio]{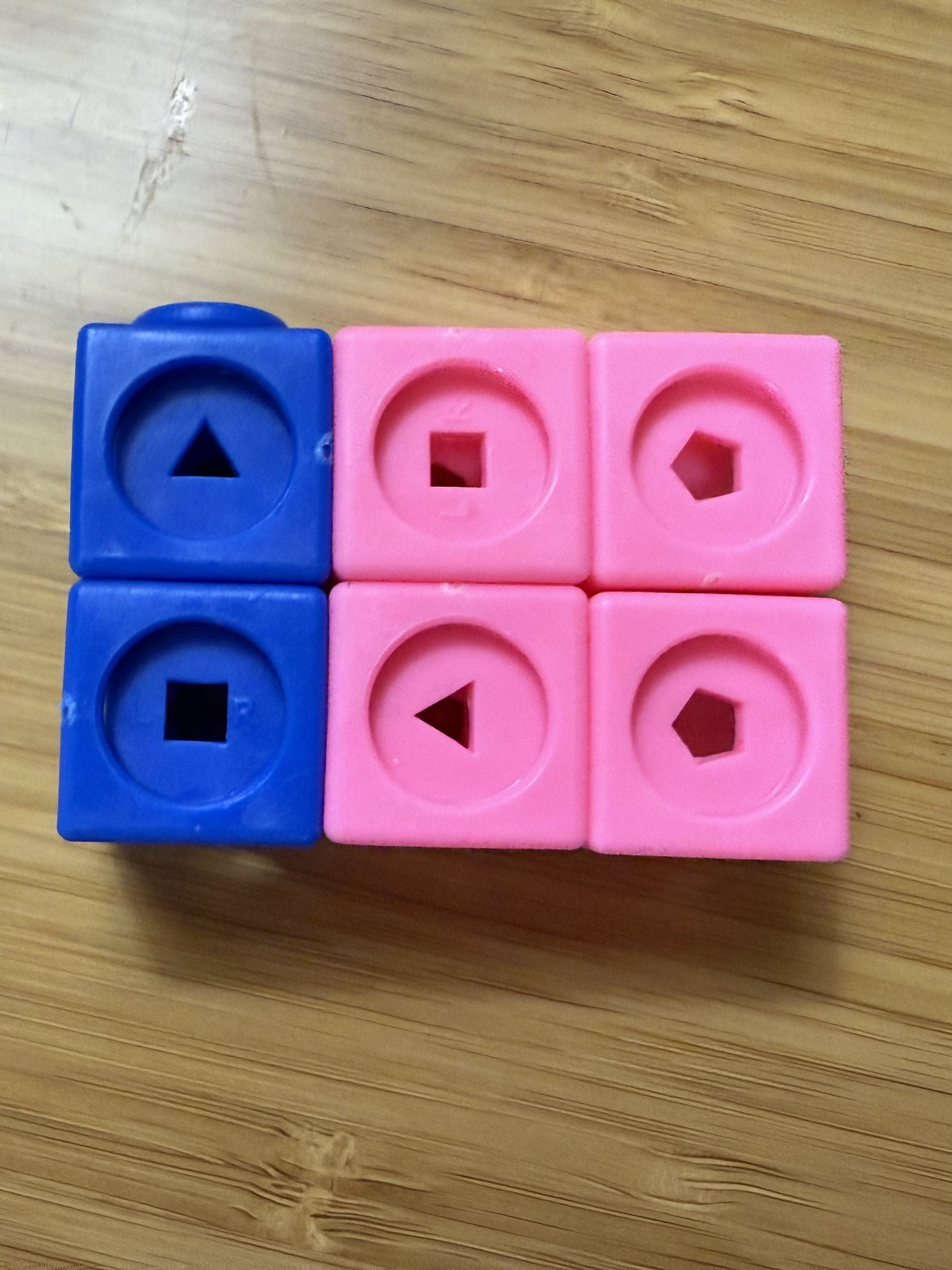}

}

\subcaption{\label{fig-8-3}Group by and summarize according to
Example~\ref{exm-summary} B.}

\end{minipage}%

\caption{\label{fig-8}Mathlink cubes summarized according to
Example~\ref{exm-summary}. Summarizing introduces a new set of mathlink
cubes in a different color than the ones used for the original data set;
in this example, we use pink.}

\end{figure}%

\subsection{Coding activity}\label{coding-activity}

After completing the coding activity with students, the subsequent
lesson can apply the exact same concepts in R using \texttt{dplyr} code.
There are several excellent publicly available examples of such lessons;
in our case we used a lesson that combined Data Science in a Box's
``Grammar of Data Wrangling'' (31) with Andrew Heiss's ``Transform data
with dplyr'' lesson (32). The lesson began with reminding students the
short hand for each of the data wrangling tasks, which happens to
exactly map to the \texttt{dplyr} syntax: \texttt{filter()},
\texttt{select()}, \texttt{mutate()}, \texttt{arrange()},
\texttt{group\_by()}, \texttt{summarize()}. We then followed the same
pattern as Section~\ref{sec-hands-on}, requesting that they complete
data analysis task, followed by a whole-class discussion of the results.
Example slides are included in the Supplemental Materials.

\subsection{Student feedback}\label{sec-data}

This activity was used in an introduction to regression and data science
course at Wake Forest University. This is a foundational course designed
to teach students data analysis through statistical computing in R.
After completing this lesson, students were invited to complete a short
evaluation. The use of this data was approved by Wake Forest
University's Institutional Review Board (IRB00025437). There were 30
students enrolled in this course; 23 completed the evaluation of this
lesson. Students were first asked to rate the statement ``How useful did
you find the''hands-on'' data manipulation lesson?'' on a scale from 1
(not very useful) to 5 (very useful). 18 students (78\%) indicated the
activity was useful, with a score of 4 or 5, and the remaining 5
students (22\%) responded neutrally, with a score of 3. They were then
asked, ``Which of the following best represents how you feel
about''hands-on'' lessons?'' 20 students (87\%) responded ``I would
prefer to see more hands on lessons in the class like the data
manipulation lesson'', while only 3 students responded ``I would prefer
to see fewer hands on lessons in class''. Finally we asked if students
had any other comments to share; two examples of positive comments were
``The physical hands-on lesson was very beneficial'', and ``The colors
and shapes really help convey the idea of data having characteristics in
a simple way.'' Only one student provided negative feedback about the
lesson, stating, ``I felt like I understand it better just doing it on
Rstudio, and following your example.''

\section{Discussion}\label{discussion}

This study demonstrates an innovative pedagogical approach to
introducing data wrangling by integrating mathlink cubes as a physical
representation of a data set. This allows students to wrap their heads
around what is ordinarily introduced in the abstract, while also
facilitating a collaborative and interactive learning environment. In
addition to providing an activity that can be incorporated in the
classroom, we have also demonstrated that this activity can be useful
and age-appropriate for undergraduate students, as evidenced by the
positive survey results reported in Section~\ref{sec-data}.

Incorporating the hand-on portion of this activity used one additional
75-minute class period that previously covered other material. In the
case of the particular course this was trialed in, this time was made up
across several ``lab'' days, where there was less time for the students
to work on their own, and more time covering the material replaced by
this additional lesson. Based on the student feedback, this hands-on
lesson was a worthwhile addition.

During implementation, I observed several key learning moments. I find
that understanding the difference between filtering (on rows) and
selecting (on columns) has confused my students in the past; during this
hands-on lesson, students also had this issue the first time they tried
filtering when we regrouped to the ``sharing'' portion of the exercise,
however since we were taking time to, as a class, notice this issue, it
seems they grasped it more quickly than in previous years when I simply
noted the potential confusion on a slide but they did not spend the time
making the mistake (or not) themselves. I noticed that the collaborative
nature of handling cubes led to productive discussions about data
transformation logic when we translated what we did with the mathlink
cubes to coding in the subsequent lesson. Additionally, I witnessed
students teaching and learning from each other while working in their
groups of three. Next time, I plan to include images of the mathlink
cubes in the subsequent coding lesson to tie the concepts more
concretely to the hands-on lesson they completed during the previous
session.

While the activity consumed additional class time, students appeared to
grasp data wrangling concepts more quickly in subsequent coding
exercises (although this was not formally assessed). The physical nature
of the activity encouraged collaboration and hopefully reduced the
anxiety some students experience when learning to code. Incorporating
opportunities such as this for students to work together on something
that is different from what they typically have seen in other statistics
classes seems to have made the class more interactive in general,
although again this was not formally assessed.

Our case study was limited to a single class of 30 students. We did not
formally assess whether this method improved learning outcomes compared
to traditional instruction. Future research could examine learning
outcomes or student performance in subsequent data analysis tasks. For
large classes, instructors could organize students into groups and
designate ``cube managers'' to handle efficient distribution. Although
this study focused on undergraduate students, these tools and activities
could potentially be adapted for both high school and graduate level
education. Additional research is needed to determine their
effectiveness across different academic levels. While we used R's
\texttt{dplyr} package, the physical concepts translate well to other
programming languages like SQL or Python.

Our findings contribute to existing literature on hands-on learning in
technical disciplines. The mathlink cube approach validates theoretical
frameworks suggesting that concrete manipulatives can bridge conceptual
understanding in complex computational domains. By transforming abstract
data transformation processes into physical interactions with mathlink
cubes, we provide a mechanism to implement these learning strategies in
an undergraduate classroom.

Our lesson offers a promising alternative to traditional abstract data
manipulation instruction, providing a tangible, interactive learning
methodology that may enhance students' conceptual understanding and
technical skill acquisition.

\section{Acknowledgements}\label{acknowledgements}

We would like to thank Ciaran Evans for a helpful discussion that led to
the successful implementation of this activity. Funding was provided by
Wake Forest University's Course Enhancement Funds.

\section{Supplemental Materials}\label{supplemental-materials}

Slides used to introduce the hands-on activity can be found here:
\url{https://docs.google.com/presentation/d/1v43TOEo9UAJPn-pdYfwmprKStQ-h2dOF9-sZdCROgVw/}

Slides used to introduce the coding activity can be found here:
\url{https://sta-112-s24.github.io/slides/04-data-manipulation#/title-slide}

\section{Data Availability}\label{data-availability}

All of the data are reported in Section~\ref{sec-data}.

\phantomsection\label{refs}

\end{document}